\documentclass{iopjournal}
\usepackage{newtxtext,newtxmath}
\usepackage{microtype}
\usepackage{amsmath}
\usepackage{bm}
\usepackage{physics}
\usepackage{siunitx}
\usepackage{graphicx}
\usepackage[linesnumbered,ruled,vlined]{algorithm2e}
\DontPrintSemicolon
\usepackage{booktabs}
\usepackage{cleveref}
\crefname{equation}{Eq.}{Eqs.}
\crefname{figure}{Fig.}{Figs.}
\crefname{table}{Table}{Tables}
\crefname{algorithm}{Algorithm}{Algorithms}

\newcommand{\br}[1]{\left(#1\right)}
\newcommand{\brs}[1]{\left[#1\right]}
\newcommand{\brc}[1]{\left\{#1\right\}}
\newcommand{\RR}{\mathbb{R}}

\newcommand{\etal}[1]{#1 \emph{et al.}}
\newcommand{\ee}{\mathcal{E}}
\newcommand{\EE}{\bm{\mathcal{E}}}
\newcommand{\hh}{\mathcal{H}}
\newcommand{\HH}{\bm{\mathcal{H}}}
\newcommand{\JJ}{\bm{\mathcal{J}}}
\newcommand{\jj}{\jmath}
\renewcommand{\Re}{\operatorname{\mathfrak{R}}}
\renewcommand{\Im}{\operatorname{\mathfrak{I}}}
\newcommand{\rnum}[1]{\text{\uppercase\expandafter{\romannumeral #1\relax}}}

\usepackage[backend=biber]{biblatex}
\begin{document}
  \articletype{Paper}
  \title{Multilayer model for coatings with arbitrary layers for superconducting radio-frequency applications}

  \author{Aaron Gobeyn$^1$\orcid{0009-0006-0431-0833}, Wolfgang Ackermann$^1$, Herbert De Gersem$^1$}
  
  \affil{$^1$Institute of Accelerator Science and Electromagnetic Fields (TEMF), TU Darmstadt}

  \email{aaron.gobeyn@tu-darmstadt.de}

    
  \begin{abstract}
    We extend the multilayer model of \etal{Kubo} for superconductor-insulator-superconductor (SIS) structures in two ways: 
    first, by generalizing it to arbitrary sequences of layers of arbitrary type, i.e. superconducting, normal conducting, and 
    insulating; and second, by accounting for all contributions, including ohmic losses and dielectric effects. We examine the maximum 
    applicable field for $(\text{SI})^n\text{S}$ structures. We find that the optimum configuration corresponds to the 
    $n=1$ case. However, the thickness of the superconducting coating layers can be reduced to below their penetration depth 
    with minor performance penalty. We discuss the ability to model transitions in SS bilayers by introducing a set of virtual layers 
    that represent the transition region through interpolated parameters. We find degradation of the maximum applicable field 
    with thicker transition layers, and a larger effective penetration depth of the electromagnetic fields. Furthermore, 
    the surface impedance of the multilayer structure is calculated using the Leontovich boundary condition, yielding a formulation 
    suitable for integration into finite element simulations. Additionally, the Poynting theorem is used to determine the loss 
    contributions of individual layers.
  \end{abstract}

  \section{Introduction}
    
  Current superconducting radio-frequency (SRF) applications, such as cavities used in particle accelerators, are primarily manufactured 
  from bulk niobium (Nb), as it has the highest critical temperature $T_c$ and highest lower critical field $H_{\text{c1}}$ among 
  pure metals \cite{Padamsee_2008}. However, it has been observed that the advantages of superconducting materials are realized only 
  within a thin surface layer ($< \SI{1}{\micro\metre}$) \cite{Valente_Feliciano_2016}. This observation has motivated the concept 
  of coated cavities with thin superconducting layers with improved material parameters. Such approaches have already been 
  successfully implemented at CERN facilities in the form of Nb/Cu cavities \cite{Calatroni_2006}. Further 
  developments such as $\text{Nb}_3\text{Sn}$ coatings on Cu substrates are under investigation \cite{Fonnesu_2026}. An extension 
  of this concept was proposed by Gurevich \cite{Gurevich_2006} in the form of superconductor-insulator-superconductor (SIS) 
  multilayer structures. In this approach, the bulk superconducting substrate is coated with a thin insulating layer, followed 
  by a thin superconducting layer with higher $T_c$ and $H_{\text{c1}}$. The superconducting thin film shields the bulk material 
  from the electromagnetic fields and attenuates them sufficiently to maintain the superconducting state at accelerating fields 
  exceeding those achievable with the bulk material alone. The insulating layer plays an important role in preventing vortex 
  penetration and suppressing Josephson effects \cite{Kubo_2016}.
  
  The multilayer model developed by \etal{Kubo} \cite{Kubo_2013,Kubo_2014} describes the electromagnetic fields penetrating the 
  multilayer structure as a function of the layer thicknesses and their material parameters. In particular, the model enables the 
  determination of the maximum applicable magnetic field, which can be used to optimize the choice of layer thicknesses. In the 
  present work, we extend this model in two respects. First, the formulation is generalized to an arbitrary number of layers of 
  arbitrary type, i.e. superconducting, normal conducting and insulating. These formulas allow the analysis of e.g. 
  SS bilayers, superconducting coatings on normal conductors, extended SIS structures of the form $(\text{SI})^n\text{S}$, 
  oxide layers on any of the preceding structures, etc. Second, we retain the contributions of normal electrons, i.e. ohmic 
  losses, and dielectric losses in our formulation. This becomes particularly relevant when determining the surface 
  impedance of the structure for applications in finite element (FE) simulations.
   
  \section{Multilayer model} \label{sec:model}
    
  \begin{figure}[!htbp]
    \centering
    \includegraphics[width=0.5\linewidth]{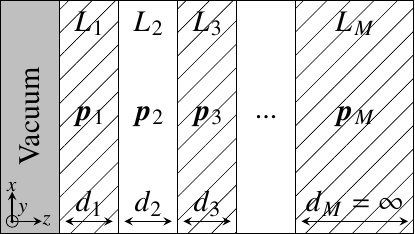}
    \caption{Multilayer structure of the cavity boundary. Each layer is labeled $L_k$ for $k = 1,\dotsc,M$ and has material 
    parameters $\bm{p}_k$ and thickness $d_k$. The first layer $L_1$ is at the inside of the cavity. The final layer $L_M$ is the
    bulk substrate.}
    \label{fig:multilayer}
  \end{figure}

  The multilayer structure is defined as illustrated in \cref{fig:multilayer}. We consider a configuration consisting 
  of $M > 1$ layers, where each layer is denoted $L_k$ for $k = 1,\dotsc,M$. Each layer $L_k$ is characterized by a vector 
  of material parameters $\bm{p}_k$, and by a thickness $d_k$. The thickness of the final layer is assumed to be effectively infinite. For 
  example, in a SIS structure, the substrate thickness is typically around \SI{2}{\milli\metre}, whereas the penetration 
  depth is only on the order of \SIrange{100}{1000}{\nano\metre}. Consequently, the fields decay well before reaching the end of the 
  substrate. 
  
  Within each layer, we assume homogeneous, isotropic, linear and frequency-dependent materials. Consequently, the 
  constitutive relations are $\bm{D} = \varepsilon \bm{E}$ and $\bm{B} = \mu \bm{H}$, where $\bm{D}$ is the displacement field, 
  $\varepsilon$ the permittivity, $\bm{E}$ the electric field strength, $\bm{B}$ the magnetic flux density, $\mu$ the permeability 
  and $\bm{H}$ the magnetic field strength. Furthermore, we consider time-harmonic fields,
    \begin{equation}
      \bm{E}(\bm{r}, t) = \Re\brs{\EE (\bm{r}) e^{- \jj \omega t}}, 
      \quad \bm{H}(\bm{r}, t) = \Re\brs{\HH (\bm{r}) e^{-\jj \omega t}},
      \quad \bm{J}(\bm{r}, t) = \Re\brs{\JJ (\bm{r}) e^{-\jj \omega t}}.
    \end{equation}
  where $\omega$ denotes the angular frequency and $\bm{J}$ is the current density. In the interior of each layer, 
  the time-harmonic Maxwell equations must be satisfied. Assuming the absence of external charges, they are given by \cite{Jackson_1999}
    \begin{equation}
      \begin{aligned}
        &\div \HH = 0, &&\curl \EE = \jj \omega \mu \HH, \\
        &\div \EE = 0, &&\curl \HH = \JJ - \jj \omega \varepsilon \EE
      \end{aligned}
    \end{equation}
  Furthermore, the interface conditions must hold at each interface between layer $L_{k-1}$ and $L_{k}$ for $k = 2, \dotsc, M$
    \begin{equation}
      \label{eqn:interface_conditions}
      \begin{aligned}
        &\bm{n} \times \br{ \EE^{(k)} - \EE^{(k-1)} } = 0, &&\bm{n} \times \br{ \HH^{(k)} - \HH^{(k-1)} } = \bm{K}_k \\
        &\bm{n} \cdot \br{ \varepsilon_k \EE^{(k)} - \varepsilon_{k-1} \EE^{(k-1)} } = \Sigma_k, &&\bm{n} \cdot \br{ \mu_k \HH^{(k)} - \mu_{k-1} \HH^{(k-1)} } = 0
      \end{aligned}
    \end{equation}
  where $\EE^{(k)}$ and $\HH^{(k)}$ denote the fields in layer $L_k$ for $k = 1, \dotsc, M$, $\bm{K}_k$ are the surface current densities, 
  $\Sigma_k$ are the surface charge densities, and $\bm{n}$ is a normal vector that points from $L_{k-1}$ to $L_k$, in the presented setup $\bm{n} = \bm{e}_z$.
  All currents are modelled within the layers, even when flowing in a very thin region. Hence the surface current densities $\bm{K}_k$ are zero. 
  Additionally, we utilize the modified Ohm's law arising from the two-fluid model \cite{London_1940}
    \begin{equation}
      \JJ = \sigma(\omega) \EE, \quad \sigma(\omega) = \sigma_n + \frac{\jj}{\omega \mu \lambda^2}
    \end{equation}
  where $\sigma_n$ is the normal conductivity and $\lambda$ is the London penetration depth. The optical conductivity 
  from the two-fluid model may be replaced by more advanced formulations, for example those derived from BCS theory 
  \cite{Mattis_1958,Nam_1967}, or more recent development on Dynes superconductors \cite{Herman_2021}. From the
  above relations, one obtains
    \begin{equation}
      \label{eqn:helmholtz}
      \laplacian \EE + \alpha^2 \EE = 0, \quad \laplacian \HH + \alpha^2 \HH = 0, 
    \end{equation}
  with 
    \begin{equation}
      \alpha(\bm{p}, \omega) = \sqrt{ \mu \varepsilon \omega^2 - \frac{1}{\lambda^2} + \jj \omega \mu \sigma_n }
    \end{equation}
  For superconducting materials, the parameter vector is given by $\bm{p} = (\mu, \varepsilon, \sigma_n, \lambda)$. In the case 
  of a normal conductor, the limit $\lambda \to \infty$ applies, reducing the number of parameters to three. For insulating materials, 
  the same limit holds; however, losses are typically not described through the normal conductivity $\sigma_n$. Instead, the 
  electric permittivity is treated as a complex quantity $\tilde{\varepsilon} = \varepsilon + \jj (\sigma_n / \omega)$, and the 
  dielectric losses are specified through the loss tangent $\tan \delta = \sigma_n / (\omega \varepsilon)$.

\begin{figure}[htbp]
    \centering
    \includegraphics[width=\linewidth]{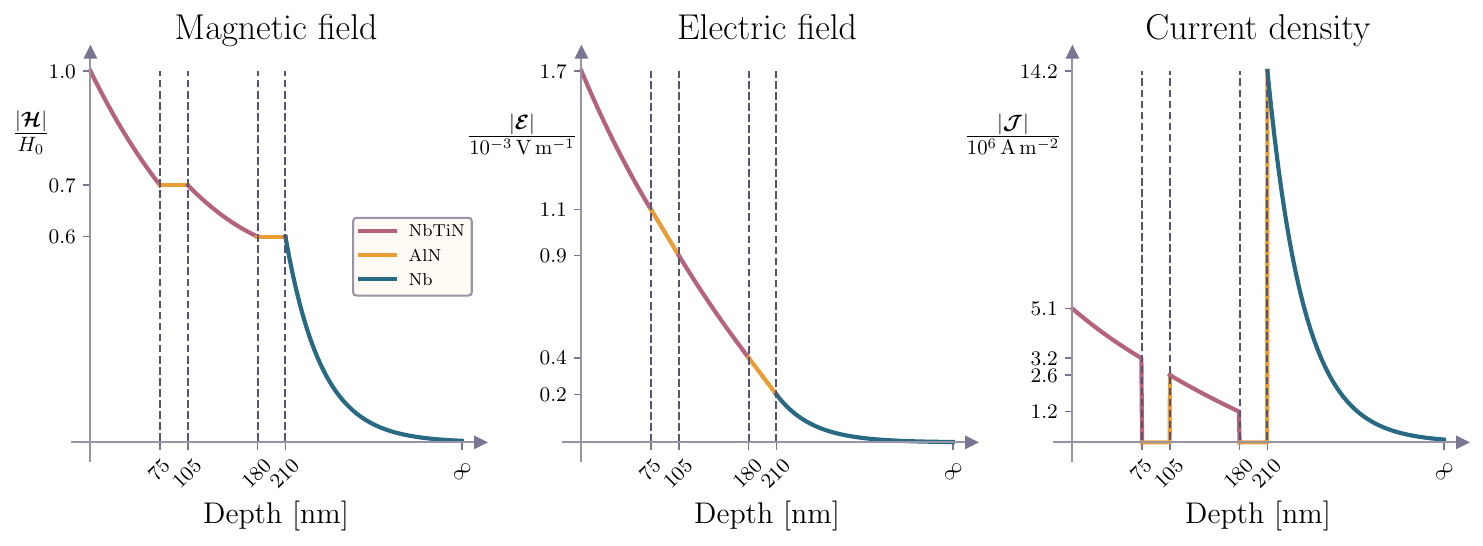}
    \caption{Electromagnetic fields and current density inside a NbTiN-AlN-NbTiN-AlN-Nb multilayer structure. The 
    material parameters used are listed in \cref{tab:parameters}. The thicknesses used for the layers are, 
    $d_1 = d_3 = \SI{75}{\nano\metre}$ and $d_2 = d_4 = \SI{30}{\nano\metre}$.}
    \label{fig:fields}
  \end{figure}

  Assuming that the penetration depth is small compared to the wavelengths of interest, the electromagnetic fields can be 
  considered uniform in the planar directions, i.e. $\partial_x \EE = \partial_y \EE = 0$. 
  The problem therefore becomes one-dimensional, i.e. $\EE(\bm{r}) = \EE(z)$. Furthermore, since $\div \EE = 0$, it follows that $\partial_z \ee_z = 0$, implying that 
  $\ee_z$ is constant. Because a non-zero constant would correspond to infinite field energy, $\ee_z = 0$. Finally, assuming linear 
  polarization, the coordinate system may be chosen such that $\ee_y = 0$, leaving $\ee_x$ as the only non-zero component of the electric field.
  Under these assumptions, the Helmholtz equation \cref{eqn:helmholtz} reduces to the ordinary differential equation
    \begin{equation}
      \dv[2]{\ee_x}{z} + \alpha^2 \ee_x = 0 \implies \ee_x(z) = A e^{\jj \alpha z} + B e^{- \jj \alpha z}
    \end{equation}
  From Faraday's law it follows that the magnetic field has $\hh_y(z)$ as the only non-zero component which is given by
    \begin{equation}
      \hh_y(z) = \frac{\alpha}{\omega \mu} \brs{A e^{\jj \alpha z} - B e^{-\jj \alpha z}}
    \end{equation}
  These solutions apply within each layer. Accordingly, we denote the fields in layer $L_k$ by $\ee_x^{(k)}(z)$ and 
  $\hh_y^{(k)}(z)$ for $k=1,\dotsc,M$, each associated with unknown coefficients $A_k$ and $B_k$. We express the 
  equations in vector form \footnote{$\alpha_k:=\alpha(\bm{p}_k,\omega)$}
    \begin{equation}
      \label{eqn:fields}
      \ee_x^{(k)} (z) = \bm{C}_k^T \bm{u}_k(z), \quad \hh_x^{(k)}(z) = \frac{\alpha_k}{\omega \mu_k} \bm{C}_k^T \bm{v}_k(z)
    \end{equation}
  with 
    \begin{equation}
      \bm{C}_k = \begin{pmatrix} A_k \\ B_k \end{pmatrix}, \quad 
      \bm{u}_k(z) = \begin{pmatrix} e^{\jj \alpha_k z} \\ e^{-\jj \alpha_k z} \end{pmatrix}, \quad 
      \bm{v}_k(z) = \begin{pmatrix} e^{\jj \alpha_k z} \\ -e^{-\jj \alpha_k z} \end{pmatrix}
    \end{equation}
  For a structure containing $M$ layers, there are $2M$ unknown coefficients that must be determined. These coefficients can be 
  obtained by imposing the interface conditions at each of the layer interfaces. The position of the interface between adjacent 
  layers $L_{k-1}$ and $L_k$ is denoted by $\zeta_k$ for $k = 2, \dotsc, M$. The special case $\zeta_1$ corresponds to the 
  boundary between vacuum and the first layer $L_1$. We finally fix the coordinate system by setting $z = \zeta_1 = 0$. From the 
  interface conditions \cref{eqn:interface_conditions} and the solutions \cref{eqn:fields} one finds that the fields 
  must be continuous over the interfaces. Considering the magnetic field at the $\zeta_1$ interface we obtain the 
  first boundary condition. Denoting the magnetic field magnitude with
  $H_0$ one obtains
    \begin{equation}
      \label{eqn:vacuum_condition}
      \hh_y^{(1)}(\zeta_1) = H_0
    \end{equation}
  Only one independent condition arises here, since the electric field magnitude is fixed by Maxwell's equations once $H_0$ is specified. 
  At each interface $\zeta_{k+1}$ between layers $L_k$ and $L_{k+1}$ for $k = 1,\dotsc,M-1$, continuity of both electric and 
  magnetic fields yields 
    \begin{align}
      \label{eqn:continuity_conditions}
      &\ee_x^{(k)} (\zeta_{k+1}) = \ee_x^{(k+1)}(\zeta_{k+1}), \\
      &\hh_y^{(k)} (\zeta_{k+1}) = \hh_y^{(k+1)}(\zeta_{k+1}), \\
    \end{align}
  Finally, the fields must vanish at infinity. As in the first condition, this yields only one independent constraint
    \begin{equation}
      \label{eqn:infty_condition}
      \lim_{z \to \infty} \hh_y^{(M)} = 0
    \end{equation}
  Substituting \cref{eqn:fields} into \cref{eqn:vacuum_condition} gives
    \begin{equation}
      \label{eqn:coeff_vacuum}
      A_1 - B_1 = \frac{\omega \mu_1}{\alpha_1} H_0
    \end{equation} 
  Similarly, applying the continuity conditions \cref{eqn:continuity_conditions} for $k = 1, \dotsc, M-1$ 
    \begin{align}
      \label{eqn:aux1}
      &\bm{C}_k^T \bm{u}_k(\zeta_{k+1}) = \bm{C}_{k+1}^T \bm{u}_{k+1}(\zeta_{k+1}), \\
      \label{eqn:aux2}
      &\bm{C}_k^T \bm{v}_k(\zeta_{k+1}) = \beta_k \bm{C}_{k+1}^T \bm{v}_{k+1}(\zeta_{k+1})
    \end{align}
  where 
    \begin{equation}
      \beta_k = \frac{\mu_k}{\mu_{k+1}} \frac{\alpha_{k+1}}{\alpha_k}
    \end{equation}
  By adding \cref{eqn:aux1,eqn:aux2} and multiplying with $(1/2)\exp(-\jj \alpha_k \zeta_{k+1})$ one obtains
    \begin{equation}
      A_k = \frac{1}{2} e^{-\jj \alpha_k \zeta_{k+1}} \bm{C}_{k+1}^T \brs{ \bm{u}_{k+1}(\zeta_{k+1}) 
      + \beta_k \bm{v}_{k+1}(\zeta_{k+1}) }
    \end{equation}
  Similarly, subtracting \cref{eqn:aux2} from \cref{eqn:aux1} and multiplying with $(1/2)\exp(\jj \alpha_k \zeta_{k+1})$ gives 
    \begin{equation}
      B_k = \frac{1}{2} e^{\jj \alpha_k \zeta_{k+1}} \bm{C}_{k+1}^T \brs{ \bm{u}_{k+1}(\zeta_{k+1}) - \beta_k \bm{v}_{k+1}(\zeta_{k+1})} 
    \end{equation}
  These equations can be written compactly as the matrix equation
    \begin{equation}
      \label{eqn:characterstic_equation}
      \bm{C}_k = \bm{\tau}_{k,k+1} \bm{C}_{k+1}
    \end{equation}
  where the characteristic matrix $\bm{\tau}_{k,k+1}$ is given by
    \begin{equation}
      \label{eqn:characterstic_matrix}
      \bm{\tau}_{k,k+1} = \frac{1}{2}
      \begin{pmatrix}
        (1+\beta_k) \exp(\jj \zeta_{k+1} \brs{-\alpha_k + \alpha_{k+1}}) & 
        (1-\beta_k) \exp(\jj \zeta_{k+1} \brs{-\alpha_k - \alpha_{k+1}}) \\
        (1-\beta_k) \exp(\jj \zeta_{k+1} \brs{\alpha_k + \alpha_{k+1}}) & 
        (1+\beta_k) \exp(\jj \zeta_{k+1} \brs{\alpha_k - \alpha_{k+1}}) \\
      \end{pmatrix}
    \end{equation}
  \cref{eqn:characterstic_equation} can essentially be viewed as a recursion relation, i.e. to know the coefficients of layer $L_k$, 
  one must know the coefficients of layer $L_{k+1}$. The recursion ends at the last layer, whose coefficients we determine by 
  substituting \cref{eqn:fields} into \cref{eqn:infty_condition}
    \begin{equation}
      \lim_{z \to \infty} \brs{A_M e^{\jj \alpha_M z} - B_M e^{-\jj \alpha_M z}} = 0
    \end{equation}
  For this limit to hold, the coefficient multiplying the exponentially diverging term must vanish. If 
  $\Im(\alpha_M) < 0$,  then $\Re(\jj \alpha_M z) > 0$ implying $A_M = 0$. Conversely, if $\Im(\alpha_M) > 0$, 
  then $\Re(-\jj \alpha_M z) > 0$ and therefore $B_M = 0$. For a superconducting substrate with typical parameters 
  $\omega{\sim}\SI{e10}{\per\second}$, $\lambda{\sim}\SI{e-7}{\metre}$ and $\sigma_n{\sim}\SI{e7}{\siemens\per\metre}$ one finds 
  $\alpha_M{\sim}(\num{e3} + \jj \num{e7})\unit{\per\metre}$. Similarly, for a normal conducting substrate, where 
  the London penetration depth does not contribute, one finds $\alpha_M{\sim}\num{e5}(1+\jj)\unit{\per\metre}$. In both 
  cases $\Im(\alpha_M) \gg 0$, leading to the conclusion that $B_M = 0$ and hence
    \begin{equation}
      \label{eqn:coeff_infty}
      \bm{C}_M = A_M \begin{pmatrix} 1 \\ 0 \end{pmatrix}
    \end{equation}
  The coefficients are then determined by propagating this conditions through the recursion relation \cref{eqn:characterstic_equation}. 
  In particular first layer coefficients $\bm{C}_1$ is determined from the final coefficients $\bm{C}_M$ by
    \begin{equation}
      \bm{C}_1 = \brs{ \prod_{k = 1}^{M-1} \tau_{k,k+1} } \bm{C}_M := \bm{T} \bm{C}_M,
    \end{equation}
  where the ordered product of all characteristic matrices is denoted by $\bm{T}$ and referred to as the transfer matrix. Using 
  \cref{eqn:coeff_infty} one obtains
    \begin{equation}
      \label{eqn:coeff_first_layer}
      A_1 = A_M T_{00}, \quad B_1 = A_M T_{10}
    \end{equation}
  where $T_{00}$ and $T_{10}$ are components of $\bm{T}$. Combining these expressions with \cref{eqn:coeff_vacuum} yields
    \begin{equation}
      A_M = \frac{\omega \mu_1}{\alpha_1 \brs{T_{00} - T_{10}}} H_0
    \end{equation}
  This determines all coefficients through the recursion relation \cref{eqn:characterstic_equation}, and therefore completely 
  specifies the fields in \cref{eqn:fields}. The resulting electromagnetic fields and current density for an example 
  $(\text{SI})^n\text{S}$ structure are shown in \cref{fig:fields}.

  \section{Maximum applicable field}
  
  The accelerating gradients achievable in SRF cavities are primarily limited by the intrinsic field limits of Nb. With modern 
  surface-preparation techniques, peak magnetic fields of approximately \SI{150}{\milli\tesla} are routinely attained \cite{Padamsee_2008}.
  In type \rnum{2} superconductors, two critical magnetic fields exist: the lower critical field $H_{\text{c1}}$ and the 
  upper critical field $H_{\text{c2}}$. For a magnetic field $H < H_{\text{c1}}$, the material is in the Meissner state, exhibiting 
  perfect diamagnetism, expelling all magnetic flux lines. When $H > H_{\text{c2}}$, superconductivity is destroyed and the 
  material transitions to the normal conducting state. In the intermediate regime $H_{\text{c1}} < H < H_{\text{c2}}$, the 
  material enters the vortex state, where magnetic flux penetrates the superconductor in the form of quantized Abrikosov vortices 
  \cite{Tinkham_1995}.

  Although the vortex state minimizes the free energy in this regime, the transition from the Meissner state to the vortex 
  state does not necessarily occur, because the Meissner state is metastable. This metastability arises due to the Bean-Livingston 
  surface barrier, which must be overcome before vortices can penetrate the material. The barrier disappears when the magnetic 
  field reaches the superheating field $H_{\text{sh}}$, which constitutes the theoretical upper limit for practical application. 
  For bulk Nb, theoretical estimates place this limit as high as \SI{240}{\milli\tesla} \cite{Padamsee_2008}, whereas the best 
  experimentally achieved values are around \SI{200}{\milli\tesla} \cite{Watanabe_2013}. 

  In the work of \etal{Kubo} \cite{Kubo_2013, Kubo_2014, Kubo_2016}, the maximum applicable field $B_{\max}$ is evaluated for 
  both SIS multilayer structures and SS bilayers. For SIS structures, they argue that $B_{\max}$ is determined by the depairing limit 
  of the superconducting coating and the empirical field limit of the substrate material. The magnetic field which exceeds these limits 
  when reaching their respective interfaces determines the maximum field that the structure can sustain. For SS bilayers, the depairing 
  limit is instead considered for both coating and substrate. Here, we extend these considerations to a general $(\text{SI})^n\text{S}$ 
  multilayer structure. The depairing limit for each superconducting thin layer corresponds to the current density at which the 
  induced magnetic field reaches the superheating field, $J_{\text{dp}}^{(k)} = H_{\text{sh}}^{(k)} / \lambda_k$, while the empirical field limit
  of the substrate is denoted by $H_{\text{emp}}^{(M)}$. Let $L_k$ be a superconducting thin layer that is first encountered by 
  the electromagnetic field at the interface $z = \zeta_k$. We then determine the value of the applied field $H_0$ for which the 
  depairing limit in this layer is reached. Denoting the resulting value with $H_{0,\text{sh}}^{(k)}$, we obtain
    \begin{equation}
      H_{0,\text{sh}}^{(k)} = \Re \brs{
        \frac{\alpha_1}{\omega \mu_1 \sigma_k(\omega)} 
        \frac{T_{00} - T_{10}}{T_{00}^{(k)} e^{\jj \alpha_k \zeta_k} + T_{10}^{(k)} e^{-\jj\alpha_k\zeta_k}}
      } \frac{H_{\text{sh}}^{(k)}}{\lambda_k}
    \end{equation}
  where we introduce 
    \begin{equation}
      \bm{T}^{(k)} = \prod_{i=k}^{M-1} \tau_{i,i+1}.
    \end{equation}
  Similarly, we determine the value $H_0$ at which the empirical field limit is reached once the field enters the substrate, 
  denoting this value with $H_{0,\text{emp}}^{(M)}$, we find
    \begin{equation}
      H_{0,\text{emp}}^{(M)} = \Re \brs{ \frac{\mu_M}{\mu_1} \frac{\alpha_1}{\alpha_M} 
      \br{T_{00} - T_{10}} e^{-\jj \alpha_M \zeta_M} } H_{\text{emp}}^{(M)}
    \end{equation}
  The maximum applicable field $B_{\max}$ is then defined by the smallest value of $H_0$ that causes either the depairing limit 
  to be reached in any of the superconducting thin layers, or the empirical limit to be reached in the substrate, i.e.
    \begin{equation}
      \label{eqn:Bmax}
      B_{\max} = \min_k \brc{\mu_k H_{0,\text{sh}}^{(k)}, \mu_M H_{0,\text{emp}}^{(M)} }.
    \end{equation}
  Here, the index $k$ runs over all superconducting thin layers. When extending the SS bilayer configuration to an arbitrary 
  number of superconducting layers, \cref{eqn:Bmax} can still be used, except that the depairing limit is also applied to the substrate.

\begin{figure}[!htbp]
    \centering
    \includegraphics[width=0.5\linewidth]{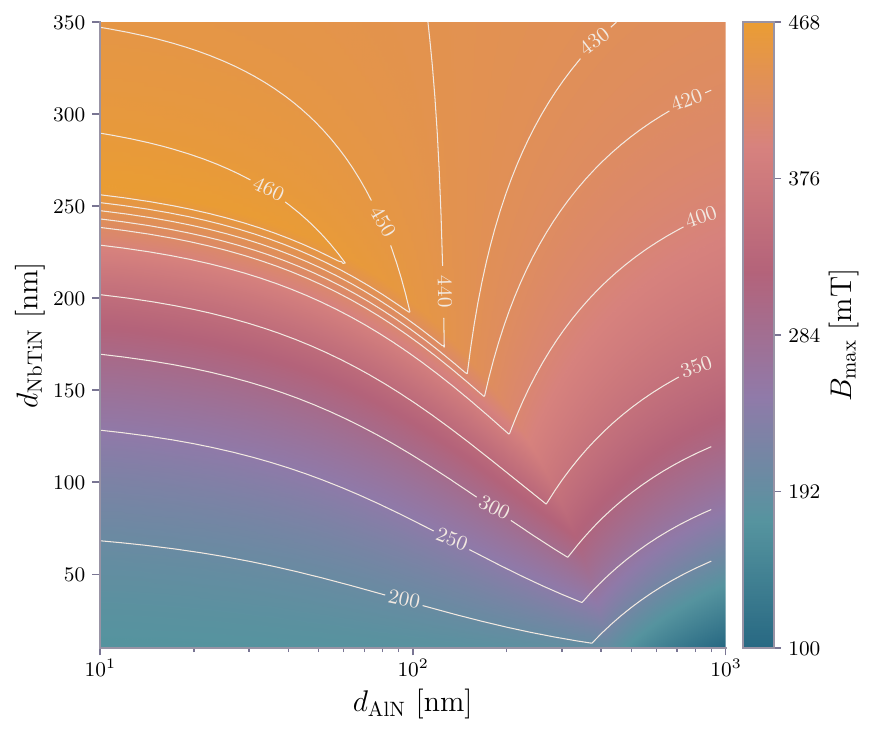}
    \caption{Maximum applicable field $B_{\max}$ of a NbTiN-AlN-Nb multilayer system as a function of the insulating 
    and coating layer thicknesses in units of \unit{\milli\tesla} at a frequency of \SI{1.3}{\giga\hertz}.
    The parameters used are listed in \cref{tab:parameters} and are temperature adjusted to $\SI{4.2}{\kelvin}$ according 
    to the two-fluid model.}
    \label{fig:bmax_sis}
  \end{figure}

  Following Ref.~\cite{Kubo_2016}, we illustrate the behavior of $B_{\max}$ using contour plots as a function of the layer 
  thicknesses. In \cref{fig:bmax_sis}, we present the results for a NbTiN-AlN-Nb structure. The optimal configuration is found 
  for layer thicknesses $d_{\text{NbTiN}} = \SI{260}{\nano\metre}$ and $d_{\text{AlN}} = \SI{10}{\nano\metre}$. Notably, the 
  optimal thickness of the NbTiN layer is approximately \num{1.4} times its penetration depth. Ideally, however, this thickness 
  should remain below the penetration depth, so that the coating does not behave as a bulk superconductor, which would 
  reintroduce vortex penetration effects that the insulating layer is intended to suppress. Furthermore, a steep decline in $B_{\max}$ 
  is observed when $d_{\text{NbTiN}}$ is reduced from its optimal value, indicating that enforcing the condition 
  $d_{\text{NbTiN}} < \lambda_{\text{NbTiN}}$ would lead to a substantial reduction in the theoretically achievable 
  performance. 

\begin{figure}[htbp]
    \centering
    \includegraphics[width=\linewidth]{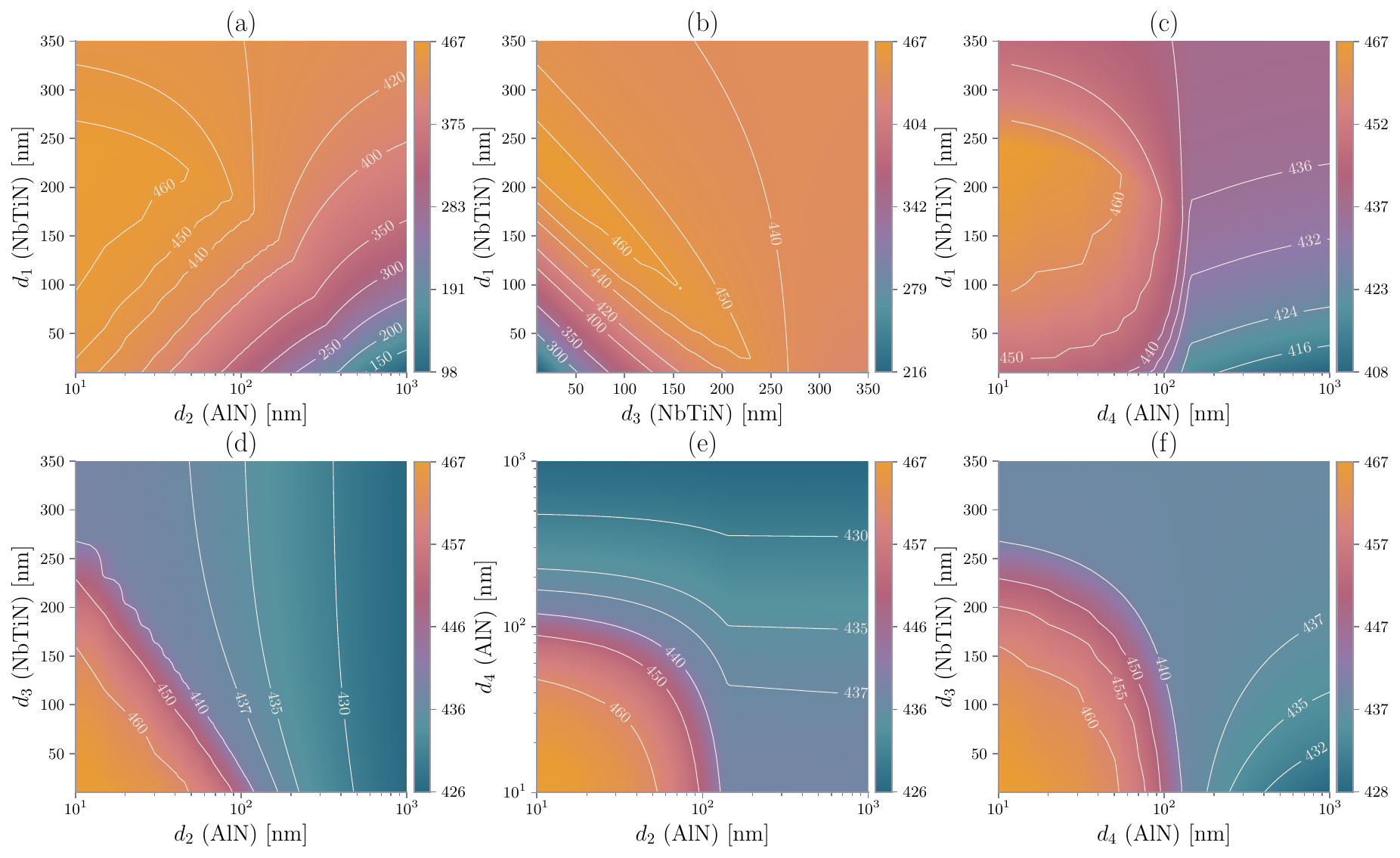}
    \caption{Maximum applicable field $B_{\max}$ of a NbTiN-AlN-NbTiN-AlN-Nb multilayer system in units of \unit{\milli\tesla} as a 
    function of the four thicknesses $d_1, d_2, d_3$ and $d_4$ of each respective non-bulk layer at a frequency 
    of \SI{1.3}{\giga\hertz}. They are displayed as six 
    contour plots, for each unique pair, by taking maximum projections along the non-varied directions. The parameters used 
    are listed in \cref{tab:parameters} and are temperature adjusted to \SI{4.2}{\kelvin} according to the two-fluid model.
    }
    \label{fig:bmax_sisis}
  \end{figure}

  We next consider the same system, but with an additional NbTiN-AlN layer pair. The corresponding results are shown in \cref{fig:bmax_sisis}, 
  where the thickness parameters not displayed in the plots are projected out by maximizing along those directions. The most 
  influential pair of layers are the first ones encountered by the field, as evidenced by the significantly larger variation 
  in $B_{\max}$ compared with other combinations (see \cref{fig:bmax_sisis}~(a)). In \cref{fig:bmax_sisis}~(d), (e) and (f), 
  similar trends are observed, with the optimal configuration approaching the lower-left corner of the parameter space where 
  corresponding layer thicknesses approach zero. Unsurprisingly then, the optimal configuration for this multilayer system is 
  given by $d_1 = d_{\text{NbTiN}}$, $d_2 = d_{\text{AlN}}$ and $d_3 = d_4 = 0$, effectively reducing the structure to the 
  three layer case from before. Nevertheless, \cref{fig:bmax_sisis}~(b) reveals a much richer parameter space when 
  considering reductions of individual layer thicknesses to below the penetration depth. For example, choosing 
  $d_1 = d_3 = \SI{150}{\nano\metre}$ and $d_2 = d_4 = \SI{10}{\nano\metre}$ yields $B_{\max} = \SI{452}{\milli\tesla}$. In this 
  configuration, the thickness of each superconducting layer remains below its respective penetration depth, while the reduction 
  in performance is only \SI{15}{\milli\tesla} relative to the optimal configuration. A similar conclusion is found 
  when the two superconducting coatings are composed of different materials, for example, when one has a higher 
  superheating field $H_{\text{sh}}$. Although the optimal configuration again eliminates one of the SI pairs, the additional 
  degrees of freedom provided by the extra layers allow for a broader range of layer thickness combinations with only minor 
  reduction in achievable $B_{\max}$.

  \begin{table*}[htbp]
    \caption{Table of material parameters for discussed multilayer structures.} 
    \label{tab:parameters}
    \centering
    \begin{tabular}{lclr}
      \toprule\toprule
      \textbf{Parameter} & \textbf{Value} & \textbf{Description} & \textbf{Source} \\
      \midrule
      $\sigma_{\text{NbTiN}}[\unit{\siemens\per\metre}]$ & \num{2.86e6} & Normal conductivity of NbTiN at \SI{4.2}{\kelvin} & Refs. \cite{Valente_Feliciano_2016,DiLeo1990} \\
      $\lambda_{\text{NbTiN}}[\unit{\nano\metre}]$ & \num{180.57} & London penetration depth of NbTiN at \SI{0}{\kelvin} & Ref. \cite{Asaduzzaman2025} \\
      $T_{c,\text{NbTiN}}[\unit{\kelvin}]$ & \num{15.4} & Critical temperature of thin layer NbTiN & Refs. \cite{GonzlezDazPalacio2023,Asaduzzaman2025}\\ 
      $B_{\text{sh},\text{NbTiN}}[\unit{\milli\tesla}]$ & \num{439} & Superheating field of NbTiN & Ref. \cite{Junginger2018IPAC} \\
      \midrule
      $\varepsilon_{r,\text{AlN}}$ & \num{10.4} & Relative permittivity of AlN & Ref. \cite{Yarar_2016} \\
      $\tan \delta_{\text{AlN}}$ & \num{2.4e-4} & Loss tangent of AlN  & Ref. \cite{Yarar_2016} \\
      \midrule
      $\sigma_{\text{Nb}}[\unit{\siemens\per\metre}]$ & \num{6.58e6} & Normal conductivity of Nb at \SI{300}{\kelvin} & Ref. \cite{rumble2017crc} \\ 
      $\text{RRR}_{\text{Nb}}$ & \num{300} & Residual resistivity ratio of Nb between \SI{300}{\kelvin} and \SI{4.2}{\kelvin}  & ... \\ 
      $\lambda_{\text{Nb}}[\unit{\nano\metre}]$ & \num{39.0} & London penetration depth of Nb at \SI{0}{\kelvin} & Ref. \cite{Maxfield1965} \\
      $T_{c,\text{Nb}}[\unit{\kelvin}]$ & \num{9.23} & Critical temperature of bulk Nb & Ref. \cite{Valente_Feliciano_2016,rumble2017crc} \\
      $B_{\text{max},\text{Nb}}[\unit{\milli\tesla}]$ & \num{170} & Maximum applicable field to bulk Nb & Ref. \cite{Kubo_2016} \\ 
      \midrule
      $\lambda_{\text{Nb}_3\text{Sn}}[\unit{\nano\metre}]$ & \num{120} & London penetration depth of $\text{Nb}_3\text{Sn}$ & Ref. \cite{Kubo_2016} \\
      $B_{\text{sh}, \text{Nb}_3\text{Sn}}[\unit{\milli\tesla}]$ & \num{454} & Superheating field of $\text{Nb}_3\text{Sn}$ & Ref. \cite{Kubo_2016} \\
      \bottomrule
    \end{tabular}
  \end{table*}

  \section{Modeling of layer transitions} 
  
  Another multilayer structure currently studied is $\text{Nb}_3\text{Sn}$ deposited on Nb or Cu substrates. Several fabrication
  approaches exist, including vapor tin diffusion on Nb substrates \cite{Posen_2021}, co-sputtering on Nb substrates \cite{Sayeed_2023}, 
  and DC magnetron sputtering on Cu substrates with Nb buffer layer \cite{Fonnesu_2026}. In all of these methods, an annealing step 
  is required to form the desired A15 crystal phase of $\text{Nb}_3\text{Sn}$, during which Sn evaporates or diffuses into the Nb. 
  As a result, the interface between the $\text{Nb}_3\text{Sn}$ layer and the Nb substrate is not sharp. The extent of the transition 
  layer is highly dependent on the deposition method, being more relevant for vapor diffusion than for sputtering-based approaches.

\begin{figure}[htbp]
    \centering
    \includegraphics[width=0.5\linewidth]{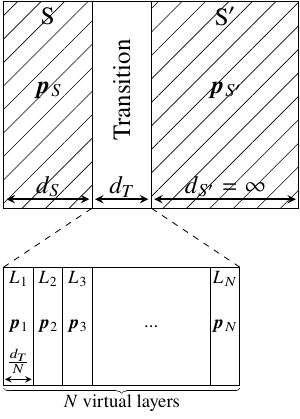}
    \caption{Graphic representation of a SS-bilayer with transition layer. The superconducting coating is denoted by S and
    has material parameters $\bm{p}_S$ and thickness $d_S$. Similarly, $\text{S}^\prime$ denotes the superconducting substrate, 
    with material parameters $\bm{p}_{S^\prime}$. The transition layer has thickness $d_S$ and is expanded into $N$ virtual 
    layers, each with thickness $d_T/N$ and parameters $\bm{p}_1, \dotsc, \bm{p}_N$. The effective structure is then of the 
    form illustrated in \cref{fig:multilayer}.}
    \label{fig:transition}
  \end{figure}

\begin{figure}[htbp]
    \centering
    \includegraphics[width=\linewidth]{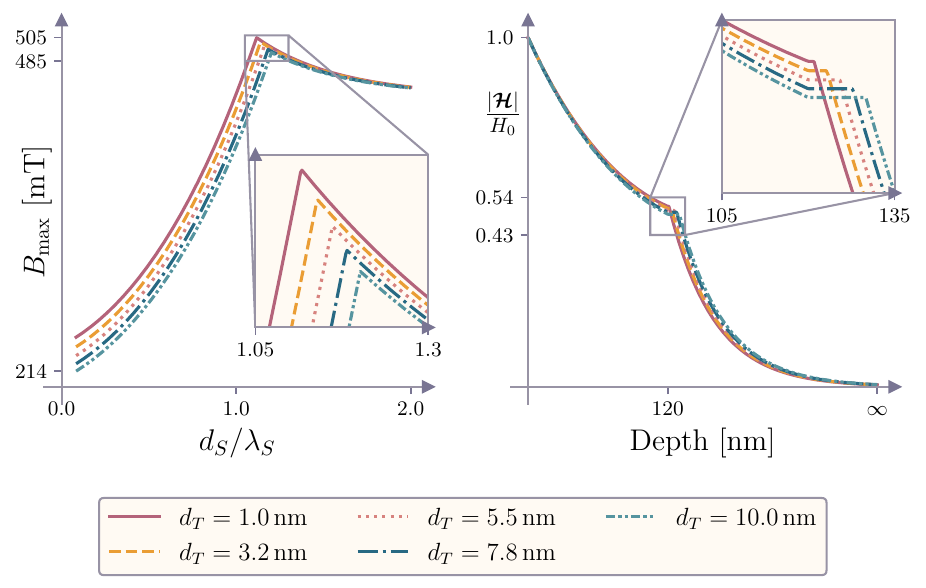}
    \caption{Maximum applicable field $B_{\max}$ (left) and magnetic field penetration (right) of a $\text{Nb}_3\text{Sn}$-Nb bilayer 
    with transition layer thicknesses ranging from \SI{1}{\nano\metre} to \SI{10}{\nano\metre}. The material parameters used are 
    listed in \cref{tab:parameters}. The transition layer is discretized into \num{100} virtual layers of equal thickness, 
    and the parameters of each virtual layer are determine through linear interpolation between the values of the substrate and the coating. 
    For the right figure, the thickness of the coating layer is equal to its penetration depth.} 
    \label{fig:bmax_transition}
  \end{figure}

  In \cref{fig:transition}, we illustrate how the multilayer model can be used to represent such a transition region. The transition 
  layer is modeled as a sequence of virtual thin superconductors, each equipped with individual material parameters. One can determine 
  these parameters through, for example, linear interpolation. The result on $B_{\max}$ and the magnetic field penetration is 
  shown in \cref{fig:bmax_transition}. The maximum applicable field $B_{\max}$ decreases by approximately \SI{10}{\milli\tesla} when 
  the transition layer increases from \SI{1}{\nano\metre} to \SI{10}{\nano\metre}. Additionally, we observe a shift in the peak, 
  indicating that slightly thicker coatings are required to reach the optimal configuration. The magnetic field distribution within 
  the material shows plateaus whose extent scales with the thickness of the transition layer. This delays the decay 
  of the magnetic field and effectively increases the penetration depth of the coating. Furthermore, we observe a slightly more rapid 
  decay within the coating when the transition layer becomes thicker. 

  It should be noted that the presented model does not account for surface roughness or grain boundaries, which may significantly 
  impact the behavior and applicability of the model. Again, this also depends on the particular deposition method in consideration, 
  where we expect it to be more applicable for vapor diffusion. Nevertheless, if the transition region can be approximated as 
  an effective intermediate layer, the presented model is still applicable.

  \section{Surface impedance}

  Determining the surface impedance of multilayer structures is particularly important for FE simulations of, for example, 
  the eigenmodes of an SRF cavity, from which the quality factor can be obtained. Because of the large discrepancy between the 
  thicknesses of the layers and the overall dimensions of the computational domain, explicitly resolving the layers within 
  the mesh is generally impractical. This limitation can be overcome by replacing the multilayer structure with an effective 
  boundary representation through the surface impedance boundary condition (SIBC). In this work, we consider only the first-order SIBC, 
  also known as the Leontovich boundary condition \cite{Landau_1960}. Higher-order SIBCs, such as the Mitzner boundary condition, 
  become relevant only when the curvature is comparable to the field attenuation \cite{Mitzner_1968}, which is not the case here. 
  For further SIBC formulation and their applications in numerical methods, see Ref. \cite{Yuferev_2018}.
  
  \subsection{Leontovich boundary condition}
   The Leontovich boundary condition reads \cite{Yuferev_2018}
    \begin{equation}
      \label{eqn:leontovich}
      \bm{n} \times \EE = Z(\omega) \brs{ \bm{n} \times \br{\bm{n} \times \HH} }
    \end{equation}
  where $\bm{n}$ is the normal vector to the boundary surface, and $Z(\omega)$ is the frequency-dependent surface impedance. 
  For the configuration shown in \cref{fig:multilayer}, the boundary corresponds to  the $z = $ plane, with the normal 
  vector $\bm{n} = \bm{e}_z$ pointing into the material. From \cref{sec:model}, we further have $\EE = \ee_x(z) \bm{e}_x$ 
  and $\HH = \hh_y(z) \bm{e}_y$. Using \cref{eqn:leontovich}, the surface impedance can therefore be expressed as
    \begin{equation}
      Z(\bm{p}, \omega) = \frac{\ee_x(z)}{\hh_y(z)} \bigg\rvert_{z = 0}
    \end{equation}
  Substituting \cref{eqn:fields} and using \cref{eqn:coeff_first_layer} yields 
    \begin{equation}
      \label{eqn:Z_leontovich}
      Z(\bm{p}, \omega) = \frac{\omega \mu_1}{\alpha_1} \frac{T_{00} + T_{10}}{T_{00} - T_{10}}
    \end{equation}
  In the special case $M=1$, the coefficients of the first and last layer coincide, implying that $\bm{T} = \text{Id}_2$, the 
  $2\times2$ identity matrix. Consequently, $Z(\omega) = (\omega \mu)/\alpha$, which reproduces the well-known expression for 
  semi-infinite normal conductors and semi-infinite superconductors within the two-fluid model. Likewise, if all layers 
  are identical, i.e. $\bm{p}_k = \bm{p}$ for all $k = 1,\dotsc,M$, then $\alpha_k = \alpha$ and $\beta_k = 1$. In this 
  case, \cref{eqn:characterstic_matrix} gives $\tau_{k,k+1} = \text{Id}_2$, and therefore $\bm{T} = \text{Id}_2$. As 
  expected, this leads to the same result as for the $M=1$ case. As an example, consider a NbTiN-AlN-Nb multilayer with material parameters 
  listed in \cref{tab:parameters} and layer thicknesses $d_{\text{NbTiN}} = \SI{260}{\nano\metre}$ and 
  $d_{\text{AlN}} = \SI{10}{\nano\metre}$. Then one finds a surface impedance $Z = \SI{1459}{\nano\ohm} - \jj \SI{1.741}{\milli\ohm}$.

  \subsection{Complex Poynting theorem}
  
  While \cref{eqn:Z_leontovich} is perfect for integration into FE simulations, it only tells us the surface impedance of the 
  entire structure, but does not distinguish the contributions of each of the layer. We therefore consider a second way 
  of determining the surface impedance which considers the contributions of each layer separately. We do so through the 
  complex Poynting theorem \cite{Jackson_1999}
    \begin{equation}
      \label{eqn:complex_poynting_thm}
      \frac{1}{2} \int_\Omega \dd V \, \br{\EE \cdot \JJ^\ast} + 2 \jj \omega \int_\Omega \dd V \, \br{w_e - w_m} 
      + \oint_{\partial \Omega} \dd S \, \br{\bm{n} \cdot \bm{\mathcal{S}}} = 0
    \end{equation}
  where 
    \begin{equation}
      w_e = \frac{\varepsilon}{4} \abs{\EE}^2, \quad w_m = \frac{\mu}{4} \abs{\HH}^2, 
      \quad \bm{\mathcal{S}} = \frac{1}{2} \br{ \EE \times \HH^\ast }.
    \end{equation}
  The quantities $w_e$ and $w_m$ are the energy densities from the electric and magnetic fields respectively, 
  and $\bm{\mathcal{S}}$ is the Poynting vector. Using the Leontovich boundary condition \cref{eqn:leontovich}, one finds that 
    \begin{equation}
      \br{\bm{n} \cdot \bm{\mathcal{S}}} \big\rvert_{z = 0} = - \frac{1}{2} \abs{H_0}^2 Z(\omega)
    \end{equation}
  Using that $\EE = \ee_x(z) \bm{e}_x$ and $\HH = \hh_y(z) \bm{e}_y$, we also find
    \begin{equation}
      \EE \cdot \JJ^\ast = \sigma^\ast(\omega) \abs{\ee_x(z)}^2, \quad 
      w_e(z) = \frac{\varepsilon^\ast}{4} \abs{ \ee_x(z) }^2, \quad 
      w_m(z) = \frac{\mu}{4} \abs{\hh_y(z)}^2
    \end{equation}
  Then, \cref{eqn:complex_poynting_thm} can be simplified and rearranged into an expression for the surface impedance as 
    \begin{equation}
      Z(\bm{p}, \omega) = \frac{1}{\abs{H_0}^2} \brs{
        \br{\sigma^\ast(\omega) + \jj \omega \varepsilon^\ast} \int_{\RR^+} \dd z \, \abs{\ee_x(z)}^2 
        - \jj \omega \mu \int_{\RR^+} \dd z \, \abs{\hh_y(z)}^2
      } 
    \end{equation}
  We split up the integrals according to the domains of each layer, i.e. 
    \begin{align}
      \label{eqn:integral_electric}
      &\int_{\RR^+} \dd z \, \abs{\ee_x(z)}^2 = \sum_{k=1}^M \int_{\zeta_k}^{\zeta_{k+1}} \dd z \, \abs{\ee_x^{(k)}(z)}^2
      := \sum_{k=1}^M I_{\ee}^{(k)}(\bm{p}_k, \omega) \\
      \label{eqn:integral_magnetic}
      &\int_{\RR^+} \dd z \, \abs{\hh_y(z)}^2 = \sum_{k=1}^M \int_{\zeta_k}^{\zeta_{k+1}} \dd z \, \abs{\hh_y^{(k)}(z)}^2
      := \sum_{k=1}^M I_{\hh}^{(k)}(\bm{p}_k, \omega)
    \end{align}
  where we define $\zeta_{M+1} = +\infty$, and $I_{\ee}^{(k)}$ denotes the integral over the electric field in layer $L_k$, 
  similarly $I_{\hh}^{(k)}$ denotes the integral over the magnetic field in layer $L_k$, for $k = 1, \dotsc, M$. Similarly, we 
  can split the surface impedance into the contributions for each layer, $Z(\bm{p}, \omega) = \sum_{k=1}^{M} Z_k(\bm{p}_k, \omega)$ where
    \begin{equation}
      Z_k(\bm{p}_k, \omega) = \frac{1}{\abs{H_0}^2} \brs{
        \br{ \sigma_k^{\ast}(\omega) + \jj \omega \varepsilon_k^\ast } I_\ee^{(k)}(\bm{p}_k, \omega) 
        - \jj \omega \mu_k I_\hh^{(k)}(\bm{p}_k, \omega)
      }
    \end{equation}
  For $k = 1,\dotsc, M-1$, the integrals \cref{eqn:integral_electric,eqn:integral_magnetic} are easily solved numerically. But for the 
  final layer $L_M$ the integration extends to infinity. Solving it analytically we find
    \begin{align}
      &I_\ee^{(M)}(\bm{p}_M, \omega) = \frac{\abs{A_M}^2}{2 \Im(\alpha_M)} e^{-2 \Im(\alpha_M) \zeta_M}, \\
      &I_\hh^{(M)}(\bm{p}_M, \omega) = \frac{1}{2 \Im(\alpha_M)} \abs{\frac{\alpha_M A_M}{\omega \mu_M}}^2 e^{-2 \Im(\alpha_M) \zeta_M}
    \end{align}
  We consider the same example as before, and find that that the contributions of each layer are: 
  $Z_{\text{NbTiN}} = \SI{649}{\nano\ohm} - \jj \SI{1.675}{\milli\ohm}$, 
  $Z_{\text{AlN}} = \SI{5e-11}{\nano\ohm} - \jj \SI{0.013}{\milli\ohm}$ and 
  $Z_{\text{Nb}} = \SI{810}{\nano\ohm} - \jj \SI{0.053}{\milli\ohm}$. As expected, and mentioned in Ref. \cite{Kubo_2016}, the 
  dielectric losses from the insulating layer are negligible compared to the other contributions. However, its contributions 
  to the reactance are not negligible. We see that a significant portion of the losses occur in the NbTiN thin layer, though most 
  losses occur in the Nb bulk substrate.

  \section{Conclusion}

  We have extended upon the model by \etal{Kubo} \cite{Kubo_2014,Kubo_2016} in two ways. We generalized the model to an arbitrary number 
  of layers of arbitrary type, i.e. superconducting, normal conducting and insulating, and accounting for all contributions, including ohmic losses 
  and dielectric effects. We have demonstrated how the maximum applicable field for $(\text{SI})^n\text{S}$ structures and 
  arbitrary number of S layers can be obtained, and shown that the optimum corresponds to the $n=1$ case. However, we have also shown 
  that $n \neq 1$ cases can be useful in order to decrease individual layer thicknesses to below their penetration depths. We have further 
  presented the ability to represent transition layers between SS-bilayers by introducing a set of virtual layers with individually 
  assignable parameters. We found that the maximum applicable field decreases as the transition layer thickness is increased, 
  and that the electromagnetic fields penetrate slightly deeper into the material. Finally, we have determined the surface impedance 
  of the structure constituting the Leontovich boundary condition so that it can be incorporated into FE simulations. 
  Additionally, we repeated this computation using the Poynting theorem such that the losses of each individual layer can be distinguished. 
  The code for performing these calculations is included in the supplementary material.

  \section*{Acknowledgements}

  The authors acknowledge financial support from the Federal Ministry of Research, Technology and Space (BMFTR), Germany, under 
  grand number 05H2024.

  \printbibliography

@article{Kubo_2016,
  title = {Multilayer coating for higher accelerating fields in superconducting
           radio-frequency cavities: a review of theoretical aspects},
  volume = {30},
  ISSN = {1361-6668},
  url = {http://dx.doi.org/10.1088/1361-6668/30/2/023001},
  DOI = {10.1088/1361-6668/30/2/023001},
  number = {2},
  journal = {Superconductor Science and Technology},
  publisher = {IOP Publishing},
  author = {Kubo, T.},
  year = {2016},
  month = dec,
  pages = {023001},
}

@book{Jackson_1999,
  author = {Jackson, J.D.},
  title = {Classical Electrodynamics},
  edition = {3},
  publisher = {Wiley},
  address = {New York},
  year = {1999},
  isbn = {978-0471309321},
}

@article{London_1940,
  title = {The high-frequency resistance of superconducting tin},
  author = {London, H.},
  volume = {176},
  ISSN = {2053-9169},
  url = {http://dx.doi.org/10.1098/rspa.1940.0105},
  DOI = {10.1098/rspa.1940.0105},
  number = {967},
  journal = {Proceedings of the Royal Society of London. Series A. Mathematical
             and Physical Sciences},
  publisher = {The Royal Society},
  year = {1940},
  month = nov,
  pages = {522–533},
}

@book{Yuferev_2018,
  title = {Surface impedance boundary conditions: a comprehensive approach},
  author = {Yuferev, S.V. and Ida, N.},
  year = {2018},
  publisher = {CRC press},
  doi = {10.1201/9781315219929 },
}

@article{Valente_Feliciano_2016,
  title = {Superconducting RF materials other than bulk niobium: a review},
  volume = {29},
  ISSN = {1361-6668},
  url = {http://dx.doi.org/10.1088/0953-2048/29/11/113002},
  DOI = {10.1088/0953-2048/29/11/113002},
  number = {11},
  journal = {Superconductor Science and Technology},
  publisher = {IOP Publishing},
  author = {Valente-Feliciano, A.-M.},
  year = {2016},
  month = sep,
  pages = {113002},
}

@article{DiLeo1990,
  title = {Niboium-titanium nitride thin films for superconducting rf
           accelerator cavities},
  volume = {78},
  ISSN = {1573-7357},
  url = {http://dx.doi.org/10.1007/BF00682108},
  DOI = {10.1007/bf00682108},
  number = {1–2},
  journal = {Journal of Low Temperature Physics},
  publisher = {Springer Science and Business Media LLC},
  author = {Di Leo, R. and Nigro, A. and Nobile, G. and Vaglio, R.},
  year = {1990},
  month = jan,
  pages = {41–50},
}

@article{Asaduzzaman2025,
  title = {Superconducting properties of thin film Nb1-xTixN studied via the NMR
           of implanted 8Li},
  volume = {37},
  ISSN = {1361-648X},
  url = {http://dx.doi.org/10.1088/1361-648X/ae0287},
  DOI = {10.1088/1361-648x/ae0287},
  number = {39},
  journal = {Journal of Physics: Condensed Matter},
  publisher = {IOP Publishing},
  author = {Asaduzzaman, M. and McFadden, Ryan M L and Thoeng, Edward and
            Kalboussi, Yasmine and Curci, Ivana and Proslier, Thomas and Dunsiger
            , Sarah R and Andrew MacFarlane, W and Morris, Gerald D and Li,
            Ruohong and Ticknor, John O and Laxdal, Robert E and Junginger,
            Tobias},
  year = {2025},
  month = sep,
  pages = {395701},
}

@article{GonzlezDazPalacio2023,
  title = {Thermal annealing of superconducting niobium titanium nitride thin
           films deposited by plasma-enhanced atomic layer deposition},
  volume = {134},
  ISSN = {1089-7550},
  url = {http://dx.doi.org/10.1063/5.0155557},
  DOI = {10.1063/5.0155557},
  number = {3},
  journal = {Journal of Applied Physics},
  publisher = {AIP Publishing},
  author = {González Díaz-Palacio, I. and Wenskat, Marc and Deyu, Getnet Kacha
            and Hillert, Wolfgang and Blick, Robert H. and Zierold, Robert},
  year = {2023},
  month = jul,
}

@article{Junginger2018IPAC,
  doi = {10.18429/JACOW-IPAC2018-THPAL118},
  url = {http://jacow.org/ipac2018/doi/JACoW-IPAC2018-THPAL118.html},
  author = {Junginger, T. and Prokscha, Thomas and Salman, Zaher and Suter,
            Andreas and Valente-Feliciano, Anne-Marie},
  language = {en},
  title = {Critical fields of SRF materials},
  journal = {Proceedings of the 9th Int. Particle Accelerator Conf.},
  volume = {IPAC2018},
  pages = {Canada},
  publisher = {JACoW Publishing, Geneva, Switzerland},
  year = {2018},
  copyright = {CC 3.0},
}

@book{rumble2017crc,
  title = {CRC handbook of chemistry and physics},
  author = {David, R.L.},
  edition = {Internet Version 2005},
  year = {2005},
  publisher = {CRC press},
  address = {Boca Raton, FL.},
}

@article{Maxfield1965,
  title = {Superconducting penetration depth of niobium},
  volume = {139},
  ISSN = {0031-899X},
  url = {http://dx.doi.org/10.1103/PhysRev.139.A1515},
  DOI = {10.1103/physrev.139.a1515},
  number = {5A},
  journal = {Physical Review},
  publisher = {American Physical Society (APS)},
  author = {Maxfield, B. W. and McLean, W. L.},
  year = {1965},
  month = aug,
  pages = {A1515–A1522},
}

@article{Yarar_2016,
  title = {Low temperature aluminum nitride thin films for sensory applications},
  volume = {6},
  ISSN = {2158-3226},
  url = {http://dx.doi.org/10.1063/1.4959895},
  DOI = {10.1063/1.4959895},
  number = {7},
  journal = {AIP Advances},
  publisher = {AIP Publishing},
  author = {Yarar, E. and Hrkac, V. and Zamponi, C. and Piorra, A. and Kienle,
            L. and Quandt, E.},
  year = {2016},
  month = jul,
}

@article{Herman_2021,
  title = {Microwave response of superconductors that obey local electrodynamics
           },
  volume = {104},
  ISSN = {2469-9969},
  url = {http://dx.doi.org/10.1103/PhysRevB.104.094519},
  DOI = {10.1103/physrevb.104.094519},
  number = {9},
  journal = {Physical Review B},
  publisher = {American Physical Society (APS)},
  author = {Herman, F. and Hlubina, R.},
  year = {2021},
  month = sep,
}

@article{Nam_1967,
  title = {Theory of electromagnetic properties of superconducting and normal
           systems. I},
  volume = {156},
  ISSN = {0031-899X},
  url = {http://dx.doi.org/10.1103/PhysRev.156.470},
  DOI = {10.1103/physrev.156.470},
  number = {2},
  journal = {Physical Review},
  publisher = {American Physical Society (APS)},
  author = {Nam, S.B.},
  year = {1967},
  month = apr,
  pages = {470–486},
}

@article{Mattis_1958,
  title = {Theory of the anomalous skin effect in normal and superconducting
           metals},
  volume = {111},
  ISSN = {0031-899X},
  url = {http://dx.doi.org/10.1103/PhysRev.111.412},
  DOI = {10.1103/physrev.111.412},
  number = {2},
  journal = {Physical Review},
  publisher = {American Physical Society (APS)},
  author = {Mattis, D. C. and Bardeen, J.},
  year = {1958},
  month = jul,
  pages = {412–417},
}

@article{Gurevich_2006,
  title = {Enhancement of rf breakdown field of superconductors by multilayer
           coating},
  volume = {88},
  ISSN = {1077-3118},
  url = {http://dx.doi.org/10.1063/1.2162264},
  DOI = {10.1063/1.2162264},
  number = {1},
  journal = {Applied Physics Letters},
  publisher = {AIP Publishing},
  author = {Gurevich, A.},
  year = {2006},
  month = jan,
}

@book{Padamsee_2008,
  title = {RF superconductivity, science, technology and applications},
  author = {Padamsee, H.},
  month = {apr},
  year = {2009},
  publisher = {John Wiley \& Sons},
  address = {New York, NY},
  isbn = {978-3-527-40572-5},
}

@article{Calatroni_2006,
  title = {20 Years of experience with the Nb/Cu technology for superconducting
           cavities and perspectives for future developments},
  volume = {441},
  ISSN = {0921-4534},
  url = {http://dx.doi.org/10.1016/j.physc.2006.03.044},
  DOI = {10.1016/j.physc.2006.03.044},
  number = {1–2},
  journal = {Physica C: Superconductivity},
  publisher = {Elsevier BV},
  author = {Calatroni, S.},
  year = {2006},
  month = jul,
  pages = {95–101},
}

@article{Fonnesu_2026,
  title = {Recipe optimization and SRF test of Cu-compatible Nb3Sn films by DC
           magnetron sputtering from a stoichiometric target},
  volume = {16},
  ISSN = {2045-2322},
  url = {http://dx.doi.org/10.1038/s41598-025-33547-w},
  DOI = {10.1038/s41598-025-33547-w},
  number = {1},
  journal = {Scientific Reports},
  publisher = {Springer Science and Business Media LLC},
  author = {Fonnesu, D. and Ford, D. and Chyhyrynets, E. and Keckert, S. and
            Knobloch, J. and Kugeler, O. and Lazzari, M. and Marconato, G. and
            Salmaso, A. and Zubtsovskii, A. and Pira, C.},
  year = {2026},
  month = jan,
}

@inproceedings{Kubo_2013,
  author = {Kubo, T., and Iwashita, Y. and Saeki, T.},
  title = {RF field-attenuation formulae for the multilayer coating model},
  booktitle = {Proceedings of the 4th International Particle Accelerator
               Conference (IPAC 2013)},
  address = {Shangai, China},
  year = {2013},
  pages = {2343-2345},
  isbn = {978-3095450-122-9},
}

@article{Kubo_2014,
  title = {Radio-frequency electromagnetic field and vortex penetration in
           multilayered superconductors},
  volume = {104},
  ISSN = {1077-3118},
  url = {http://dx.doi.org/10.1063/1.4862892},
  DOI = {10.1063/1.4862892},
  number = {3},
  journal = {Applied Physics Letters},
  publisher = {AIP Publishing},
  author = {Kubo, T. and Iwashita, Yoshihisa and Saeki, Takayuki},
  year = {2014},
  month = jan,
}

@book{Landau_1960,
  title = "Electrodynamics of continuous media",
  author = "Landau, L.D. and Lifshitz, E.M",
  publisher = "Pergamon Press",
  year = 1960,
  address = "Oxford, England",
}

@article{Mitzner_1968,
  title = {Effective boundary conditions for reflection and transmission by an
           absorbing shell of arbitrary shape},
  volume = {16},
  ISSN = {0096-1973},
  url = {http://dx.doi.org/10.1109/TAP.1968.1139283},
  DOI = {10.1109/tap.1968.1139283},
  number = {6},
  journal = {IEEE Transactions on Antennas and Propagation},
  publisher = {Institute of Electrical and Electronics Engineers (IEEE)},
  author = {Mitzner, K.},
  year = {1968},
  month = nov,
  pages = {706–712},
}

@book{Tinkham_1995,
  title = "Introduction to superconductivity",
  author = "Tinkham, M.",
  publisher = "McGraw-Hill",
  edition = 2,
  month = nov,
  year = 1995,
  address = "New York, NY",
  isbn = {978-0-070-64878-4},
}

@article{Watanabe_2013,
  title = {Development of the superconducting rf 2-cell cavity for cERL injector
           at KEK},
  volume = {714},
  ISSN = {0168-9002},
  url = {http://dx.doi.org/10.1016/j.nima.2013.02.035},
  DOI = {10.1016/j.nima.2013.02.035},
  journal = {Nuclear Instruments and Methods in Physics Research Section A:
             Accelerators, Spectrometers, Detectors and Associated Equipment},
  publisher = {Elsevier BV},
  author = {Watanabe, K. and Noguchi, S. and Kako, E. and Umemori, K. and
            Shishido, T.},
  year = {2013},
  month = jun,
  pages = {67–82},
}

@article{Sayeed_2023,
  title = {Fabrication of superconducting Nb3Sn film by Co-sputtering},
  volume = {212},
  ISSN = {0042-207X},
  url = {http://dx.doi.org/10.1016/j.vacuum.2023.112019},
  DOI = {10.1016/j.vacuum.2023.112019},
  journal = {Vacuum},
  publisher = {Elsevier BV},
  author = {Sayeed, N. and Pudasaini, Uttar and Eremeev, Grigory V. and
            Elsayed-Ali, Hani E.},
  year = {2023},
  month = jun,
  pages = {112019},
}

@article{Posen_2021,
  title = {Advances in Nb3Sn superconducting radiofrequency cavities towards
           first practical accelerator applications},
  volume = {34},
  ISSN = {1361-6668},
  url = {http://dx.doi.org/10.1088/1361-6668/abc7f7},
  DOI = {10.1088/1361-6668/abc7f7},
  number = {2},
  journal = {Superconductor Science and Technology},
  publisher = {IOP Publishing},
  author = {Posen, S. and Lee, J and Seidman, D N and Romanenko, A and Tennis, B
            and Melnychuk, O S and Sergatskov, D A},
  year = {2021},
  month = jan,
  pages = {025007},
}
\end{document}